%% file: ms.tex
\documentclass[iop]{emulateapj}

\usepackage{graphicx}
\usepackage{color}
\usepackage{enumerate}

\begin{document}

\shortauthors{Tran et al.}
\shorttitle{Anticorrelated Nature of $O-C$ Curves}
\title{THE ANTICORRELATED NATURE OF THE PRIMARY AND SECONDARY
ECLIPSE TIMING VARIATIONS FOR THE KEPLER CONTACT BINARIES}

\author{K. Tran\altaffilmark{1}, A. Levine\altaffilmark{2}, S. Rappaport\altaffilmark{1}, T. Borkovits\altaffilmark{3}, Sz. Csizmadia\altaffilmark{4}, B. Kalomeni\altaffilmark{5} }  

\altaffiltext{1}{M.I.T. Department of Physics and Kavli
Institute for Astrophysics and Space Research, 70 Vassar St.,
Cambridge, MA, 02139; sar@mit.edu} 
\altaffiltext{2}{37-575 M.I.T. Kavli Institute for Astrophysics and Space Research, 70 Vassar St.,
Cambridge, MA, 02139; aml@space.mit.edu} 
\altaffiltext{3}{Baja Astronomical Observatory, H-6500 Baja, Szegedi \'ut, Kt. 766, Hungary; Konkoly Observatory, MTA CSFK, H-1121 Budapest, Konkoly Thege M. \'ut 15-17, Hungary; ELTE Gothard-Lend\"ulet Research Group, H-9700 Szombathely, Szent Imre herceg \'ut 112, Hungary; borko@electra.bajaobs.hu} 
\altaffiltext{4}{Institute of Planetary Research, German Aerospace Center, Rutherfordstrasse 2, 12489, Berlin, Germany szilard.csizmadia@dlr.de} 
\altaffiltext{5}{Department of Astronomy and Space Sciences, University of Ege, 35100 Bornova-Izmir, Turkey; Department of Physics, Izmir Institute of Technology, Turkey}

\begin{abstract}
We report on a study of eclipse timing variations in contact binary systems,  using long-cadence lightcurves in the {\em Kepler} archive. As a first step, `observed minus calculated' ($O-C$) curves were produced for both the primary and secondary eclipses of some 2000 {\em Kepler} binaries.  We find $\sim$390 short-period binaries with $O-C$ curves that exhibit (i) random-walk like variations or quasi-periodicities, with typical amplitudes of $\pm$200-300 seconds, and (ii) anticorrelations between the primary and secondary eclipse timing variations.  We present a detailed analysis and results for 32 of these binaries with orbital periods in the range of $0.35 \pm 0.05$ days.  The anticorrelations observed in their $O-C$ curves cannot be explained by a model involving mass transfer, which among other things requires implausibly high rates of $\sim$$0.01 \, M_\odot {\rm yr}^{-1}$. We show that the anticorrelated behavior, the amplitude of the $O-C$ delays, and the overall random-walk like behavior can be explained by the presence of a starspot that is continuously visible around the orbit and slowly changes its longitude on timescales of weeks to months.  The quasi-periods of $\sim$$50-200$ days observed in the $O-C$ curves suggest values for $k$, the coefficient of the latitude dependence of the stellar differential rotation, of $\sim$0.003$-$0.013.
\end{abstract}


\section{Introduction}
\label{sec:intro}

Contact binary stars occur relatively frequently among binaries (Rucinski~1998). 
A contact binary system consists of two dwarf stars, most often from the
F, G, and K spectral classes, that are surrounded by a common
convective envelope.  The orbital period distribution peaks in the 8
to 12 hour range.  Most systems, though not all, have orbital periods
between 0.2 and 1.0 days (Maceroni \& van't Veer~1996; Paczy\'nski et
al.~2006). While the masses of the two component
stars of a contact binary are typically unequal, the two stars usually
have approximately equal surface temperatures due to the effects of mass and energy
transfer between the components via a common convective envelope
(Lucy~1968). The properties of the envelope, the energy
transfer between the components, and the overall internal structure of
the components have been investigated by many authors (see, e.g.,
K\"ahler~2002; Webbink 2003; K\"ahler~2004; Csizmadia \& Klagyivik~2004; Li et
al.~2004; Yakut \& Eggleton~2005; Stepie\'n \& Gazeas~2012).  Eclipsing contact 
binaries are often referred to as W UMa systems in honor of the prototype.

Some variable stars that were classified as contact binaries in
earlier studies are now considered otherwise; their light curves are
thought to merely mimic the light curves of true contact binaries. For
example, while AW UMa is actually a semi-detached system with a
material ring, it exhibits a light curve much like that of a contact
binary (Pribulla \& Rucinski~2008). This example and others
demonstrate that it may be difficult to determine in practice whether
a binary is a contact or a semi-detached system based only on a
photometric time series.

Although contact binaries make up an important part of the Galactic
stellar population, their formation and final-stage evolutionary
states are still not clear (Paczy\'nski et al.~2006; Eggleton~2006).
Possible formation processes and evolutionary outcomes have recently
been summarized by Eggleton~(2012).  Many, if not all, contact
binaries may be members of triple star systems, which could drive the
formation of these extremely close binaries through a combination of
the Kozai mechanism and tidal friction (Robertson \& Eggleton~1977;
Kozai~1962; Kiseleva et al.~1998; Fabrycky \& Tremaine~2007). It is
also thought that rapidly rotating single stars may be formed from the
coalescence of the components of contact binaries (Li et al.~2008;
Gazeas \& Stepie\'n~2008).  These open questions about the formation,
evolution, and final state of contact binaries make them one of the
most intriguing classes of objects in stellar astrophysics
(Eggleton~2012).

Many contact binaries show signs of stellar activity, presumably
because the component stars are rapid rotators with deep convective
zones. Doppler imaging has revealed that some contact binaries are
almost fully covered by rather irregular spot-like structures (AE Phe, 
Maceroni et al.~1994 and Barnes et al.~2004; YY Eri, Maceroni et al.~1994;
VW Cep, Hendry \& Mochnacki~2000; SW Lac, Senavci et al.~2011). Signs 
of high levels of coronal activity are often apparent; this helps explain why
contact binaries can be relatively strong X-ray emitters (Geske et
al.~2006). It may be noted that the first flare event which was
simultaneously observed both in X-rays and at radio wavelengths from a
star other than our Sun was from the contact binary VW Cephei (Vilhu
et al.~1988).  Ground based multicolor photometry demonstrated an
H$\alpha$ excess in two contact binary systems that is thought to have
a coronal origin, and to be related to the presence of dark spots on
the photosphere (Csizmadia et al.~2006).

Several contact binaries exhibit night-to-night light curve variations which
may be explained by fast spot evolution on orbital- (or even suborbital-) period 
timescales (Csizmadia et al.~2004). On the other hand, the eclipse timing 
variations of several contact binaries show quasi-periodic oscillations on a 
time-scale of $\sim$10 years, and those features were interpreted as indirect 
evidence of solar-like magnetic cycles (see e.g., Qian 2003; Awadalla et al. 
2004; Borkovits et al.~2005; Pop \& Vamos
2012). These results are in accord with the work of Lanza \&
Rodon\'o (1999), which concluded that the timescales for magnetic
modulations are strongly and positively correlated with the orbital periods
of the systems. This relation makes contact binaries excellent
laboratories in which to investigate the temporal variations and
evolution of stellar spots, in part because the timescales of the variations
are shorter than in other types of binary and single stars.  However, the
shortest among these timescales can be problematic to study using ground-based
observatories because they are comparable to the length of an Earth night.

Kalimeris et al.~(2002) noted that the migration of starspots on the
surface(s) of the constituent stars in short-period binaries,
especially contact binaries, could affect measurements of eclipse
times and thereby mimic changes in the orbital period.  Kalimeris et
al.~(2002) also showed that the perturbations to the observed minus
calculated ($O-C$) eclipse-time curves would generally have amplitudes
smaller than $\sim$0.01 days, and could appear to be quasiperiodic on
timescales of a few hundred days or so if the spot migration is
related to differential rotation of the host star.

The planet-finding {\em Kepler} mission (Borucki et al.~2010; Koch et
al.~2010; Caldwell et al.~2010) has observed more than 150,000 stars
over the past four years.  The monitoring of each star is nearly
continuous, and the photometric precision is exquisitely high (Jenkins
et al.~2010a; 2010b).  These capabilities have led to the discovery of
more than 2600 planet candidates (Batalha et al.~2013), and also of a
comparable number of binary stars (Slawson et al.~2011; Matijevi\v{c}
et al.~2012). Some 850 of the {\em Kepler} binaries have been classified 
as contact, overcontact\footnote{In contact systems the two stars just fill 
their respective Roche lobes while in overcontact binaries both components
overfill their Roche lobes and are surrounded by a low-density common 
envelope (see e.g., Wilson 1994).  In this work we use the terms ``contact 
binary'' and ``overcontact binary'' interchangeably, but only because the 
real differences between the two types are not material to the work we 
present in this paper.}, or elliposidal light variable (`ELV') systems 
(Slawson et al.~2011; Matijevi\v{c} et al.~2012).

In this work we report on a study of eclipse timing variations of binaries
in archival {\em Kepler} data, with a particular focus on contact and
overcontact binaries.  In Section~2 we describe the data
preparation, the estimation of the eclipse times, and the production
of $O-C$ curves for each contact binary.  In Sect.~3 we present the
$O-C$ data for an illustrative selection of 32 contact binaries (out of the
several hundred we found) which exhibit common interesting features 
including random-walk like, or quasi-periodic excursions in the $O-C$ 
behavior with amplitudes of $\sim$$\pm$ 300 sec,
and a generally anticorrelated behavior in the $O-C$ curves for the
primary and secondary eclipse minima or ellipsoidal-light-variation 
minima.  In Sect.~4 we extend the work
of Kalimeris et al.~2002 in order to explain some of these
characteristics with a very simple model involving a cool (or hot)
spot on one of the stars that drifts slowly around the star on
timescales of weeks to months.  We discuss the significance of our
results in Sect.~5.

\vspace{0.3cm}
\section{Data Analysis}
\subsection{Data Preparation}

The present study is based on {\em Kepler} long-cadence (LC)
lightcurves. To start, we retrieved the LC lightcurve files for
Quarters 1 through Quarter 13 for all of the candidates in the latest
{\em Kepler} eclipsing binary catalog (Slawson et al.~2011) that were
available at the Multimission Archive at STScI (MAST).  We used the
lightcurves made with the PDC-MAP algorithm (Stumpe et al.~2012; Smith
et al.~2012), which is intended to remove instrumental signatures from
the flux time series while retaining the bulk of the astrophysical
variations of each target. For each quarter, the flux series was
normalized to its median value.  Then, for each target, the results
from all available quarters were concatenated together in a single
file.  We also checked our results using the SAP-MAP processed
data set, and the results are unchanged.

The next step in the data processing was to apply a high-pass filter
to each stitched light curve.  A smoothed light curve was obtained by
convolving the unsmoothed light curve with a boxcar function of
duration equal to the known binary period.  The smoothed light curve
was then subtracted from the unsmoothed light curve.  This procedure
largely removes intensity components with frequencies below about half
the binary frequency, while leaving largely intact temporal structures
that are shorter than the binary period.  Periodically-recurring
features of the light curve are essentially unaffected.

\begin{figure}
\begin{center}
\includegraphics[width=0.99 \columnwidth]{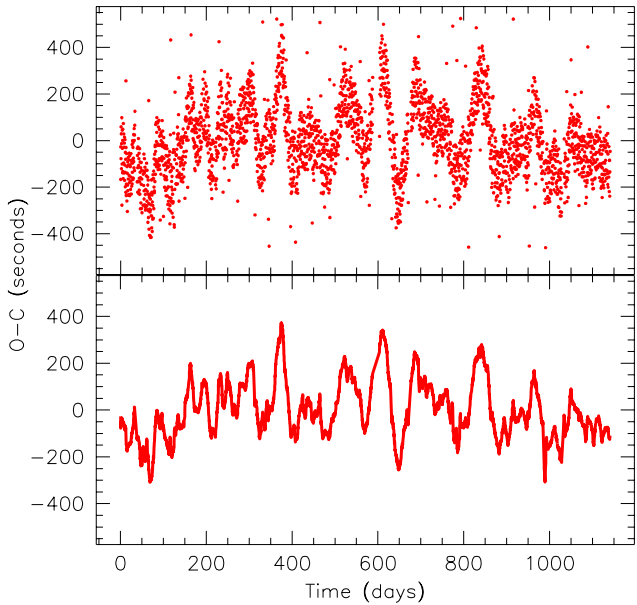}
\caption{ Illustrative $O-C$ curve for one of the contact binaries considered in this work, KIC 2159783 (``KIC'' refers to the Kepler Input Catalog; Batalha et al. 2010).  Top panel: raw $O-C$ curve.  Bottom panel: $O-C$ curve after performing a 5-day boxcar smoothing operation.  The smoothed versions of the $O-C$ curves are the ones displayed in the remainder of this work.}
\label{fig:illus_omc}
\end{center}
\end{figure}

\subsection{$O-C$ Eclipse Times}
\label{sec:ecltime}

Since long-cadence data with relatively coarse time resolution were
used for the present study, an interpolation method was needed to
estimate eclipse times with an accuracy better
than $\sim$1700 s. The algorithm utilized for the determination of
eclipse times consists simply of identifying the flux values in the
light curve that represent local minima, and then fitting a parabola
to that value and the immediately preceding and succeeding values.
The time of the minimum of the fitted parabola is used as the time of
eclipse minimum (see Rappaport et al.~2013 for details of the
algorithm).  This algorithm provides excellent accuracy for
short-duration eclipses, but loses some accuracy when the eclipse
duration is longer than $\sim$10
long-cadence samples.  For contact binaries with periods between
$\sim$0.2 and 1 day, however, the algorithm works well and typically
yields eclipse times subject to an rms scatter of $\sim$30 s.

\begin{figure*}
\begin{center}
\includegraphics[width=0.95 \textwidth]{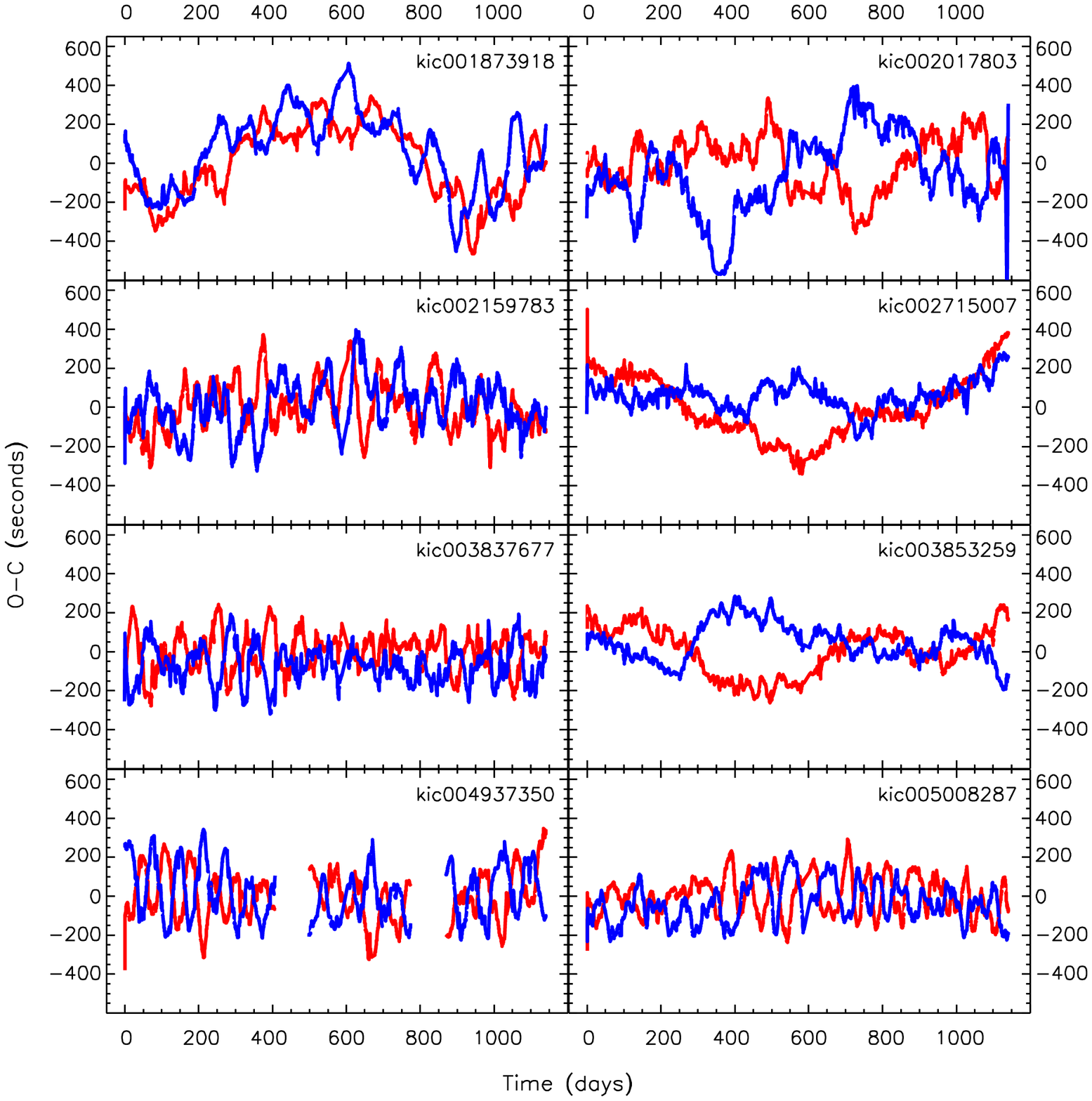}
\caption{ A sample of $O-C$ curves for an illustrative set of eight {\em Kepler} contact binary systems with KIC numbers in the range of 1873918 to 5008287.  The $O-C$ curves for the primary and secondary eclipses are shown as red and blue curves, respectively.  The curves have been smoothed over a 5-day interval, comprising typically 10-15 eclipses.  These $O-C$ curves typically exhibit random-walk-like or quasi-periodic behavior.  The $O-C$ curves for the primary and secondary eclipses are often anticorrelated. }
\label{fig:bigomc1}
\end{center}
\end{figure*}

\begin{figure*}
\begin{center}
\includegraphics[width=0.95 \textwidth]{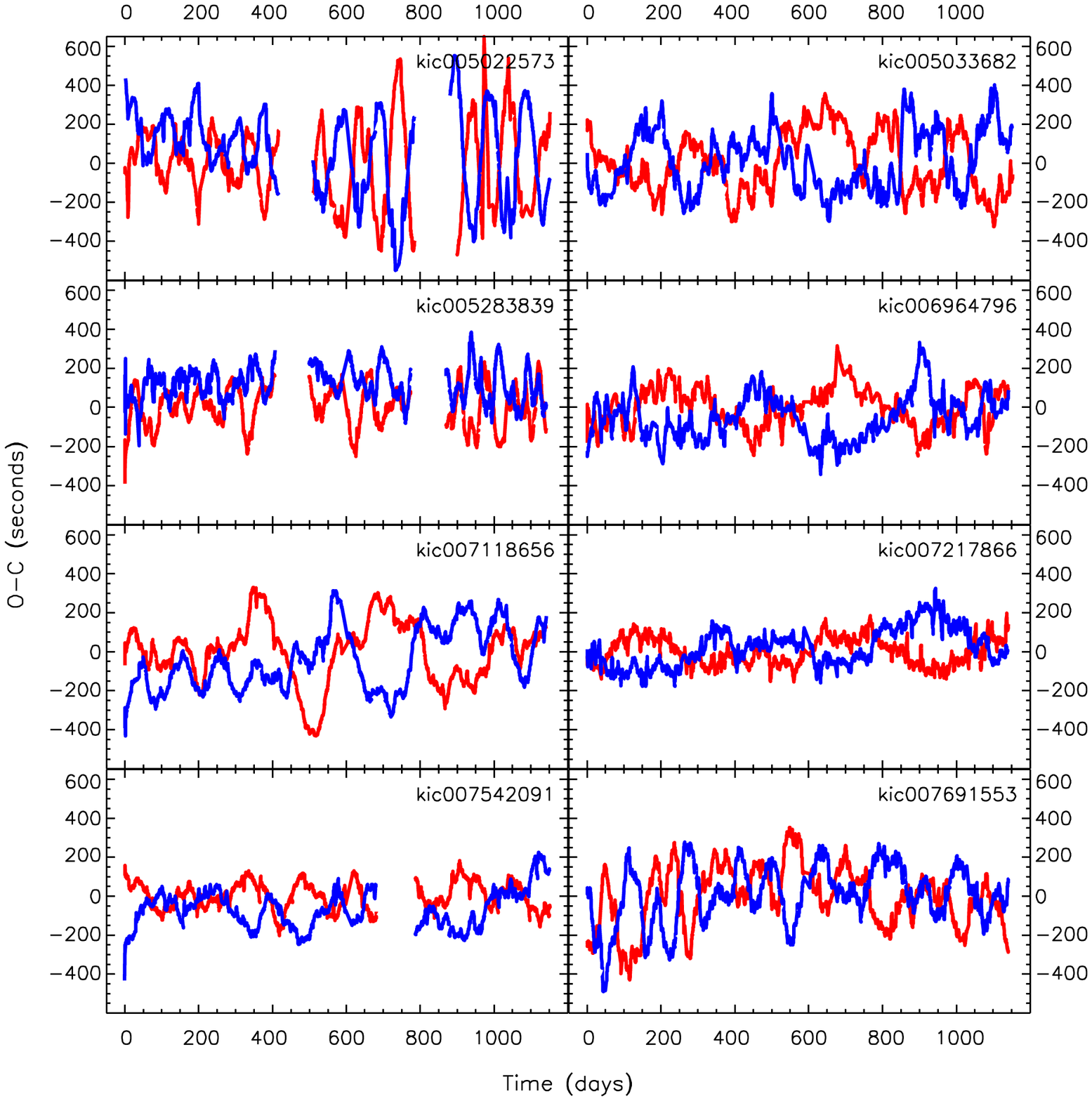}
\caption{ A sample of $O-C$ curves for an illustrative set of eight contact binary systems with KIC numbers in the range of 5022573 to 7691553.  The specifications are otherwise the same as for Fig.~\ref{fig:bigomc1}.}
\label{fig:bigomc2}
\end{center}
\end{figure*}

\begin{figure*}
\begin{center}
\includegraphics[width=0.95 \textwidth]{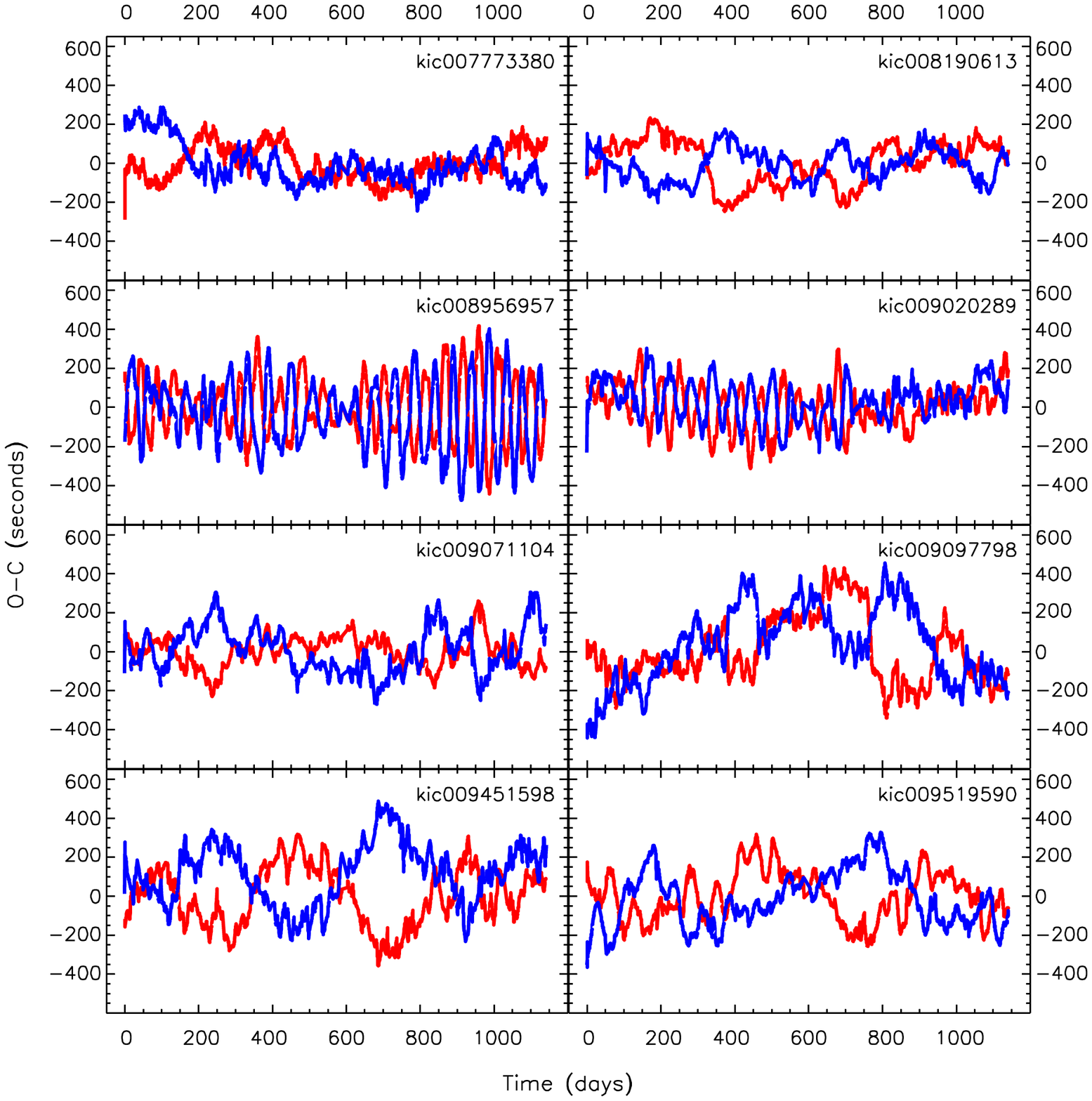}
\caption{  A sample of $O-C$ curves for an illustrative set of eight contact binary systems with KIC numbers in the range of 7773380 to 9519590.  The specifications are otherwise the same as for Fig.~\ref{fig:bigomc1}. }
\label{fig:bigomc3}
\end{center}
\end{figure*}

\begin{figure*}
\begin{center}
\includegraphics[width=0.95 \textwidth]{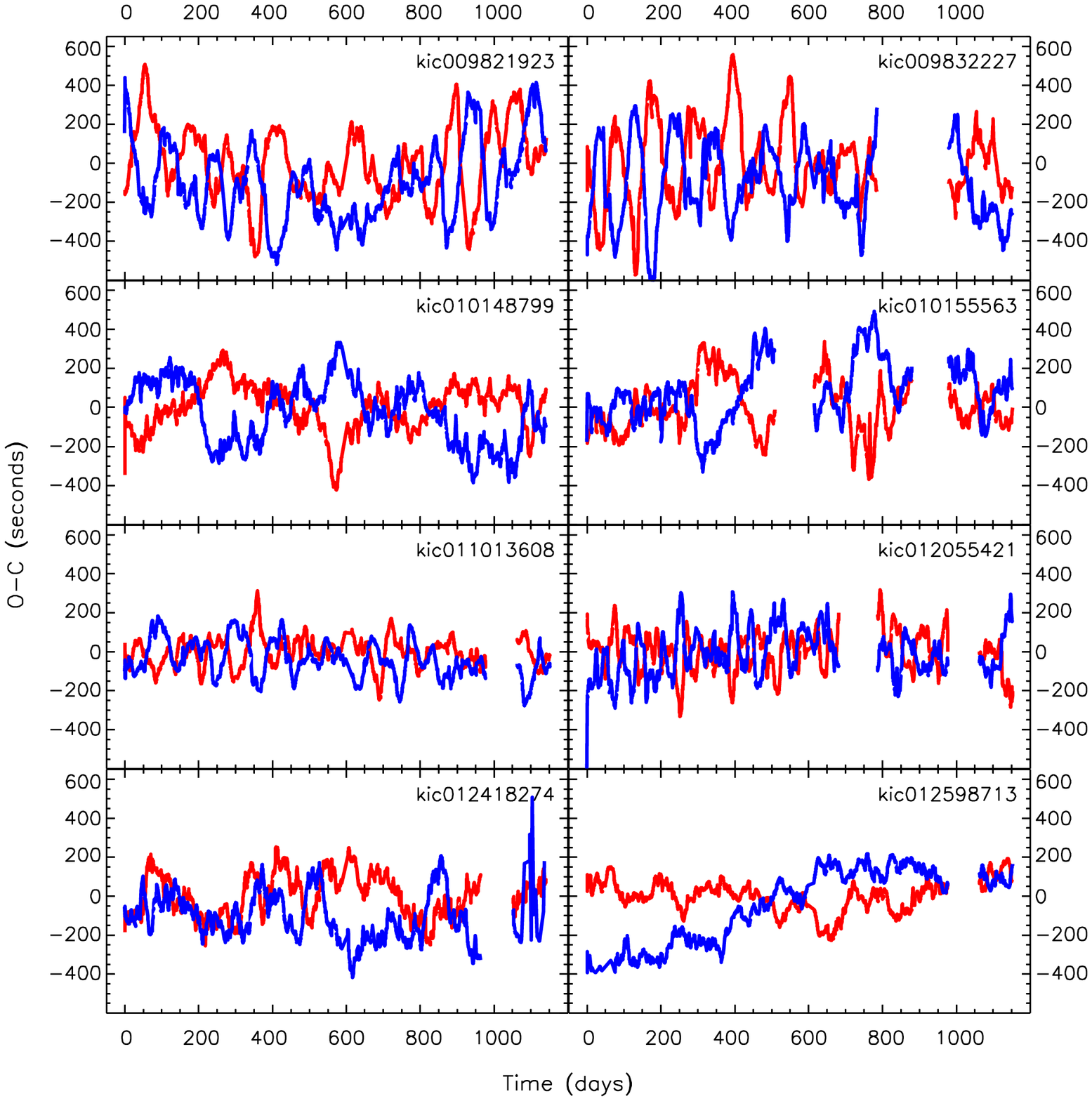}
\caption{  A sample of $O-C$ curves for an illustrative set of eight contact binary systems with KIC numbers in the range of 9821923 to 12598713.  The specifications are otherwise the same as for Fig.~\ref{fig:bigomc1}. }
\label{fig:bigomc4}
\end{center}
\end{figure*}

Once the times of the primary eclipses were found for each source, a
linear function consisting of the eclipse cycle count times the
orbital period taken from the Slawson et al.~(2011) catalog was used
in order to form an `observed minus calculated' (or ``$O-C$'') curve.
If the binary orbit is circular and is not perturbed by a third body
in the system, and if the shape of the eclipses remains constant,
the $O-C$ curve should be a straight line.

In addition to the primary eclipses, $O-C$ curves were also calculated
for the {\em secondary} eclipses of all the binary systems.

For each binary, the $O-C$ curve for the {\em primary} eclipse was fit
with a linear function to determine the best average of the orbital
period over the $\sim$3-year interval of the {\em Kepler} data set (Quarters 1-13).
This corrected period was used to produce final $O-C$ curves for {\em
both} the primary and secondary eclipses.  

Since most of the
interesting variations in the $O-C$ curves occur on time scales of
weeks to months, the $O-C$ curves were smoothed to reduce the
amplitudes of high-frequency variations without substantially
degrading longer-term features.  This procedure was accomplished by convolving
each $O-C$ curve with a boxcar function with a five day duration, thereby typically 
averaging over some $10-15$ $O-C$ values, i.e., orbital cycles (see Fig.~\ref{fig:illus_omc} 
for an illustrative sample).  This operation, of course, will remove any physical $O-C$
variability on scales $\lesssim 5$ days, but there are some artifacts, 
e.g., beats between the orbital period and the {\em Kepler} long-cadence 
integration interval of 29.4 min, that make a search for short periodicities in 
the $O-C$ curves difficult.

\section{Contact Binary $O-C$ Curves}

In the process of examining the eclipse-timing-variation $O-C$ curves for all $\sim$2000 binaries, while searching for evidence for the presence of third bodies (Rappaport et al.~2013), we discovered that a very large fraction of the shorter period binaries had a set of common features in their $O-C$ curves.  These features include (i) random-walk or quasi-periodic variations, and (ii) anticorrelated behavior between the $O-C$ curves of the primary and secondary eclipses.  We found that some 390 of the short period binaries exhibited these features in their $O-C$ curves (see Table \ref{tab:stats} for the numbers as a function of orbital period).  

We used the numerical morphological classification scheme of Matijevi\v{c} et al.~(2012) to characterize the binaries that seem to exhibit both properties listed above.  The results are displayed in Table \ref{tab:stats}.  There are some 306, 49, 30, and 4 of these binaries in the orbital period ranges $0.2-0.5$, $0.5-1$, $1-2$, and $>2$ days, respectively.  The mean, median, minimum, maximum, and standard deviation of the Matijevi\v{c} et al.~(2012) morphology parameter, $c$, are given for each orbital period category.  It is clear that the systems of interest are mostly in the orbital period range $0.2-0.5$ days, where contact and overcontact binaries are found.  The numbers fall off sharply with increasing orbital period.  

The Matijevi\v{c} et al. (2012) numerical morphological descriptor, between 0 and 1, is given for each binary. The sense of this classification scheme is that morphological values of $\lesssim $ 0.5 correspond to detached binaries, while values in the ranges $0.5-0.7$ and $0.7-0.8$ correspond to semidetached and overcontact binaries, respectively. Values higher than 0.8 are ellipsoidal light variables and ``uncertain'' classifications, and a number of these may not be eclipsing.  As can be seen from Table \ref{tab:stats}, as well as by looking at the actual distribution of values, some 2/3 of the binaries of interest have $c \gtrsim 0.8$ and $\sim$1/3 have $0.7 < c < 0.8$.  This implies that the vast majority of them are contact binaries, with a substantial fraction exhibiting largely ellipsoidal light variations (`ELV') rather than pronounced eclipses.  

For contact systems, the transition between purely ELV behavior, partial eclipses, and full eclipses is a smooth one.  Thus, we do not make a sharp distinction between eclipsing contact binaries of the W UMa class vs.~contact binaries that exhibit pure ELVs.  We therefore use the terms `eclipse times' and `times of minima', or `eclipses' and `minima', somewhat interchangeably. 

\begin{deluxetable}{l | cccc | c}
\tablewidth{0pt}
\tabletypesize{\scriptsize}
\tablecaption{\label{tab:stats} {Binary Morphology Statistics}}
\tablehead{
	\colhead{} &
	\colhead{} &
	\colhead{$P_{\rm orb}$ (days)}  & 
	\colhead{} &
	\colhead{} &
	\colhead{This} \\
	\colhead{Parameter} & 
	\colhead{$0.2-0.5$} &
	\colhead{$0.5-1$} & 
      	\colhead{$1-2$} &
	\colhead{$~~~~~>2~~~~~$} &
	\colhead{Work}
}
\startdata 
\# Systems	& 306 & 49 & 30 & 4 & 32 \\
$\langle c \rangle$	& 0.86 & 0.73 & 0.75 & 0.82 & 0.93 \\
Median $c$	& 0.87 & 0.69 & 0.78 & 0.83 & 0.91 \\ 
$\delta_c$	& 0.09 & 0.14 & 0.20 & 0.11 & 0.05 \\
$c_{\rm min}$	& 0.65 & 0.53 & 0.47 & 0.68 & 0.81 \\
$c_{\rm max}$ & 1.0 & 0.98 & 1.0 & 0.92 & 1.0 
\enddata
\tablecomments{Binary morphology statistics for the short period {\em Kepler} binaries with anticorrelated $O-C$ curves for the primary and secondary minima (\url{http://keplerebs.villanova.edu/}; Matijevi\v{c} et al.~2012).  The morphology index, $c$ is defined such that detached systems have values $\lesssim 0.5$; semi-detached systems are in the range $0.5-0.7$; and overcontact binaries lie in the range $0.7-0.8$.  Systems with morphology index above $\sim$0.8 are ellipsoidal variables and or uncertain categories. }
\end{deluxetable}

\subsection{Illustrative $O-C$ Curves for Contact Binaries}
\label{sec:CBOmC}

From the set of 390 systems whose $O-C$ curves exhibit (i) random-walk or quasi-periodic variations, and (ii) anticorrelated behavior between the primary and secondary eclipses, we selected 32 systems to display and discuss in some detail. These all fall into the period range of $0.2-0.5$ days, where the vast majority of these systems lie, but are otherwise indistinguishable from the other $\sim$275 systems that we identified in this period range.  We focus on this set for the remainder of the paper.  The morphology statistics of this group of 32 are summarized in Table \ref{tab:stats}.  

The $O-C$ curves for these 32 illustrative systems are shown in Figs. 
\ref{fig:bigomc1} -- \ref{fig:bigomc4}, and some of the system parameters
are listed in Table \ref{tab:source}.  
There are several potentially important features to note about the
selected sets of $O-C$ curves presented in this paper.  The
peak-to-peak amplitudes of the short-timescale random-walk like
behavior are typically about 500 seconds, although a few systems
exhibit somewhat higher amplitude variations.  The latter includes KIC
5022573 (top left panel of Figure~\ref{fig:bigomc2}).  The
characteristic time scales of the quasiperiodicities vary greatly, but
are often in the range of $\sim$50-200 days.

The selected $O-C$ curves also exhibit clear anticorrelated behavior
on at least some time scales and for some time intervals, even when
positive correlations between the primary and secondary curves are
evident on relatively long time scales.  For example, the $O-C$ curves
for KIC 1873918 (see Figure~\ref{fig:bigomc1}) are anticorrelated over
the 100-day time scales of their quasiperiodic variations, but show an
overall positive correlation on timescales over $\sim$800 days.

\begin{figure}
\begin{center}
\includegraphics[width=0.99 \columnwidth]{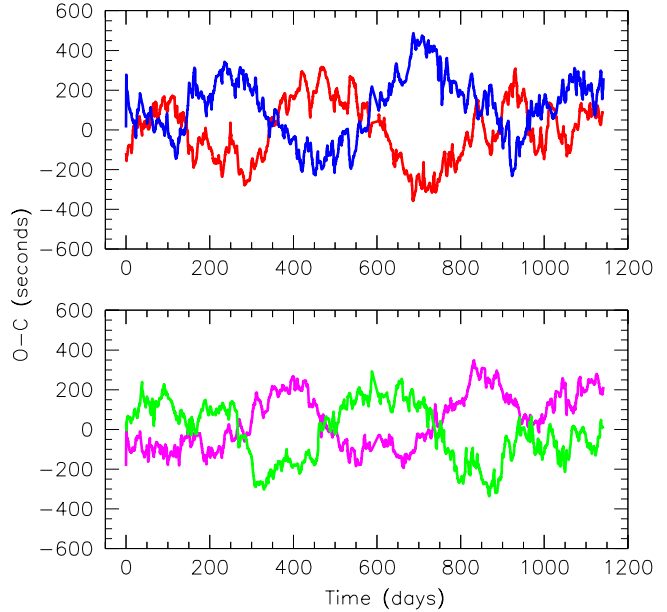}
\caption{ The O-C curves for KIC 9451598 showing anticorrelated behavior between the minima (top panel) as well as the maxima (bottom panel). Note that the curves in the bottom panel are $\sim$90$^\circ$ out of phase with respect to those in the top panel.}
\label{fig:mins_max}
\end{center}
\end{figure}

For this sample of contact binaries, $O-C$ curves were also calculated
for the times of the two {\em maxima} in each orbital cycle.  
The $O-C$ curves for the two
maxima as well as the two eclipse minima of one of the contact
binaries, KIC 9451598, are shown in Figure~\ref{fig:mins_max}. The two
$O-C$ curves for the eclipse maxima are clearly anticorrelated with
respect to each other in the same way as are the $O-C$ curves for the 
minima. In addition, the plot suggests that the $O-C$ curves
of the maxima are offset in phase from the curves of the minima by
about 90$^\circ$, i.e., that the rate of change in one curve is maximal
at the amplitude extrema in the other curve.  We attempt to explain both 
the 180$^\circ$ and 90$^\circ$ phase shifts with a simple starspot model
in Sect. 4.

\begin{figure}
\begin{center}
\includegraphics[width=0.99 \columnwidth]{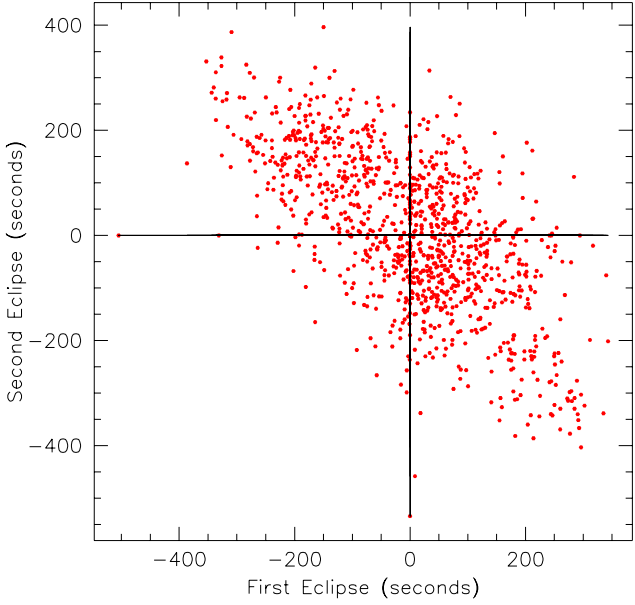}
\caption{ A point-by-point correlation plot of the binned O-C curves for the primary and secondary eclipses of KIC 7691553. The negative slope of the plot clearly demonstrates the anticorrelation of the $O-C$ curves, with a correlation coefficient of $-0.5$. }
\label{fig:crosscorr_dot}
\end{center}
\end{figure}

To demonstrate more quantitatively the anticorrelated behavior between
the $O-C$ curves of the primary and secondary minima, we show a
point by point correlation plot for one system: KIC 7691553 ($O-C$
curves shown in the bottom right panel in Fig.~\ref{fig:bigomc2}) in
Fig.~\ref{fig:crosscorr_dot}. To create this particular plot, the
$O-C$ values were averaged in one-day time bins.  The plot shows a
clear negative correlation, and confirms what is seen in a visual
inspection of the $O-C$ curves.  The formal correlation coefficient is
$-0.5$ for the particular example shown in Fig.~\ref{fig:crosscorr_dot}.
For systems that are dominated by anticorrelated $O-C$ curves, the
correlation coefficients range down to $-0.77$, with a median value of
$-0.42$.

In addition to computing and displaying correlation plots for
the contact binaries in our sample, we also computed formal cross
correlation functions (CCFs) for all the pairs of (one-day rebinned)
eclipse $O-C$ curves.  The plots mostly confirm the characteristic
timescales observed in the $O-C$ curve variations, as well as the
anticorrelated behavior at zero time lag.

Fourier transforms of the $O-C$ curves for our sample of contact
binaries do not show, in general, any strong peaks, except for the
known beat frequencies between the orbital period and the long-cadence
sampling time (see Rappaport et al.~2013 for details).  Plots of $\log
A_\nu$ vs. $\log \nu$, where $A_\nu$ is the Fourier amplitude and
$\nu$ is the frequency, generally show more or less linear relations
with logarithmic slopes of approximately $-1.0$ to $-1.3$ that are similar
to Fourier spectra associated with random-walk
behavior.

\section{Models}

\subsection{Period Changes}

Most of the structure seen in the $O-C$ curves for the contact
binaries cannot represent actual changes in the orbital periods (see
Kalimeris et al.~2002 for a related discussion).  The argument is
simple.  For circular orbits like those expected for contact binaries,
period changes would produce similar, i.e., {\em positively
correlated}, effects in both the primary and secondary $O-C$ curves.
Therefore, the variations associated with anticorrelated behavior
cannot be the result of orbital period changes.

In addition, mass transfer could not drive such rapid
changes in the $O-C$ curves even if the primary and secondary eclipses were
not anticorrelated. This may be understood quantitatively by representing a small
portion of the $O-C$ curve as:
\begin{eqnarray}
O-C \simeq \tau \sin(2 \pi t/T)
\end{eqnarray}
where $\tau$ and $T$ are rough measures of the amplitude and cycle
time of the undulations in the $O-C$ curve, and $t$ is the time. If
the variations were caused by orbital period changes, the
second derivative of the $O-C$ curve would be related to the first derivative of the
orbital period as
\begin{eqnarray}
\frac{\dot P_{\rm orb}}{P_{\rm orb}} = \frac{d^2 \null}{dt^2} (O-C) = -\frac{4 \pi^2}{T^2} \tau \sin(2 \pi t/T).
\end{eqnarray}
It is straightforward to show that mass transfer in a binary results in
$\dot P_{\rm orb}/P_{\rm orb} \approx \dot
M/M$.  Therefore, the implied mass transfer rate would be of order
\begin{eqnarray}
\frac{\dot M}{M} \approx  4 \pi^2 \frac{\tau}{T^2}.
\end{eqnarray}
For characteristic $O-C$ amplitudes of $\sim\pm$200 seconds and cycle
times of $\sim$50-200 days (see
Figs.~\ref{fig:bigomc1}-\ref{fig:bigomc4}), we find implied mass
transfer rates of $\dot M/M \simeq 0.001-0.01$ yr$^{-1}$.  These rates
are implausibly large even for a contact binary.  More physically reasonable
mass transfer rates have been inferred for contact binaries such as 
$\sim$$10^{-7} \,M_\odot$ yr$^{-1}$ in VW Boo, a typical overcontact binary 
(Liu et al.~2011), and in several W UMa systems (Borkovits et al.~2005).

\subsection{Slightly Eccentric Orbits}

The orbits of contact binaries must generally be circular or nearly
circular. However, it is perhaps conceivable that perturbations from a
third body, or even some stochastic mass-exchange, magnetic event, or
other physical process could induce a very minor eccentricity from
time to time.  The apsidal motion that would then ensue would result
in $O-C$ variations, to first order in eccentricity, with amplitudes
of $\pm (P_{\rm orb}/2\pi)2e\cos \omega $ for the primary and
secondary minima, respectively (Gimenez \& Garcia-Pelayo 1983);
$\omega$ is the longitude of periastron.  Consequently, an $O-C$
amplitude of $\pm$300 sec in a binary with an orbital period of 0.3
days requires a minimum eccentricity of $\sim$0.04.  Since this is
implausibly high for a contact binary, slightly eccentric orbits are
unlikely to be the cause of most of the anticorrelated behavior that
is apparent in the observed amplitudes.

\subsection{A Simple Starspot Model}

\subsubsection{Spot Visibility}

Here we consider a simple geometric model wherein a single spot on one
of the stars might produce the anticorrelated behavior seen in the
timing of the primary and secondary minima.  In this simplistic
picture, the stars are taken to be spherical and to be rotating
synchronously with the orbit.  The $\hat{z}$ direction is defined to
be parallel to the orbital angular momentum vector; the stars revolve
in the $x-y$ plane. The observer is located in the $y-z$ plane and
views the system with conventional orbital inclination angle $i$.  The
unit vector in the direction from the system toward the observer is
then $\vec{V} = \cos i \,\hat{z} + \sin i \,\hat{y} $. For a spot
located at colatitude $\alpha$, the angle from the stellar pole, and
stellar longitude, $\ell$, the unit vector pointing from the center of
the star through the spot is $\vec{S} = \sin \alpha \,\sin ( \omega t
+ \ell) \, \hat{x} + \sin \alpha \cos (\omega t +\ell) \, \hat{y} +
\cos \alpha \, \hat{z} $, where $\omega$ is the orbital angular
velocity.  A starspot located at $\ell = 0^\circ$ and $\alpha = 90^\circ$
is defined to lie along the line segment connecting the two stellar centers.

The spot is assumed to occupy a small portion of the surface of the
star and is taken to radiate in a Lambertian manner. The projected
area of the spot normal to the line of sight is proportional to the
cosine of the angle between the normal to the spot area and the
direction toward the observer, $\vec{V}\cdot \vec{S}$. If this dot
product is negative, the spot is on the hemisphere of the star facing
away from the observer and is not visible. When the dot product is
positive, the spot is not occulted, and if limb darkening may be
neglected, the apparent brightness of the star is changed by the
presence of the spot according to the expression
\begin{eqnarray}
\Delta F = \epsilon \left[ \cos \alpha \cos i + \sin \alpha \sin i \cos( \omega t  + \ell) \right]
\label{eqn:view_vector}
\end{eqnarray}
where $\epsilon$ is a constant with dimensions of flux assumed to be
much less than the overall flux from the binary.  In order for the
spot to remain continuously visible, $\Delta F$ must always be positive. Such
a condition requires that $\cos \alpha \cos i > \sin \alpha \sin i$,
or, equivalently, $\alpha + i <$ 90$^\circ$.

\subsubsection{Analytic Estimate of the $O-C$ Amplitudes}

When no spots are present, the light curve of a contact binary may be
represented with sufficient accuracy by
\begin{eqnarray}
\mathcal{B} = -B  \cos 2 \omega t  
\label{eqn:binary_LC}
\end{eqnarray}
where time $t=0$ corresponds to the time of primary minimum (here
indistinguishable from the secondary minimum) and where the constant
term has been dropped.  In this crude model, $B$ is the modulation
amplitude due to both ellipsoidal light variations and eclipses.

When one spot is present, the observed flux as a function of time,
again aside from constant terms, will then be:
\begin{eqnarray}
\mathcal{F} =  -B  \cos 2 \omega t  + \epsilon \sin \alpha \sin i \cos( \omega t  + \ell) .
\label{eqn:flux}
\end{eqnarray}
We can now examine analytically how the two minima and the two maxima
are shifted in phase relative to the case of no spot.  The analysis is
straightforward since it is assumed that $\epsilon \ll B$ as mentioned
above.  After expanding the relevant trigonometric functions for small
excursions about their nominal extrema, and neglecting high order
terms, the four phase shifts, in radians, are found to be:
\begin{eqnarray}
\Delta \phi_{\rm min,n} &=& -(-1)^n \, \frac{\epsilon \sin \alpha \sin i}{4 B}~\sin \ell  \label{eqn:dphi} \label{eqn:phi12} \\
\Delta \phi_{\rm max,n} &=& (-1)^n \, \frac{\epsilon \sin \alpha \sin i}{4 B}~\cos \ell \label{eqn:dphix}
\end{eqnarray} 
where $n= 1$ or 2 for the first or second, respectively, minimum or maximum. 

It is immediately clear that (i) the shifts of the times of the two
minima are anticorrelated; (ii) the shifts in the two times of maxima
are anticorrelated; and (iii) the changes in the times of the two
maxima are 90$^\circ$ out of phase with respect to the times of the
two minima.  It is also clear from these expressions that the phase
shifts depend on the spot longitude; a near uniform migration in
longitude with time leads to a quasiperiodic $O-C$ curve.

In some timing analyses, the eclipse center is defined as the midpoint
between the ingress and egress times, e.g., at their half intensity points.  
Perturbations to such eclipse centers can also be worked out analytically
using the same formalism and approximations as discussed above.  The 
corresponding shifts in the eclipse times are the same as given by eqn.~(\ref{eqn:phi12}),
except that they are larger by a factor of $\sqrt{2}$.

The (half peak-to-peak) amplitude of the $O-C$ variations seen as the spot
migrates around the star at constant $\alpha$ can be computed from eqn.~(\ref{eqn:phi12})
and is given in units of time by:
\begin{eqnarray}
 \tau = \frac{\epsilon \sin \alpha \sin i}{4 B} \frac{P_{\rm orb}}{2 \pi}
\end{eqnarray}
The coefficient quantifying the photometric strength of the spot may
be estimated by
\begin{eqnarray}
\epsilon \simeq \frac{4 \Delta T}{T} \frac{\pi r_{\rm spot}^2}{ \pi R_{\rm 1}^2} \frac{B_0}{2}
\end{eqnarray}
where $r_{\rm spot}$ is the radius of the spot, $R_{\rm 1}$ is the
radius of the star with the spot, $\Delta T$ is the decrement
(increment) in temperature of the cool (hot) spot, and $B_0$ is the
mean brightness of the binary.  Finally, if we define an eclipse depth
in terms of the fractional decrease in intensity at the minimum,
$\xi \equiv B/B_0$, the expression for the amplitude of the $O-C$
shifts becomes:
\begin{eqnarray}
\tau = \frac{1}{4 \pi \xi} \sin \alpha \sin i  \frac{ \Delta T}{T} \frac{r_{\rm spot}^2}{R_{\rm 1}^2}  P_{\rm orb} .
\label{eqn:dtau}
\end{eqnarray}
For the illustrative parameter values of $\xi = 0.04$, $\alpha =
45^\circ$, $i = 40^\circ$, $r_{\rm spot}/R_{\rm 1} = 0.2$ (equivalent
to a spot radius of 11.5$^\circ$ of arc on the stellar surface),
$\Delta T/T = 0.15$, and $P_{\rm orb} = 8.7$ hours, we find $ \tau
\sim$170 s, a value similar to those seen in the $O-C$ observations.

\subsubsection{Eclipses and Spot Occultations}

\begin{figure}
\begin{center}
\includegraphics[width=0.99 \columnwidth]{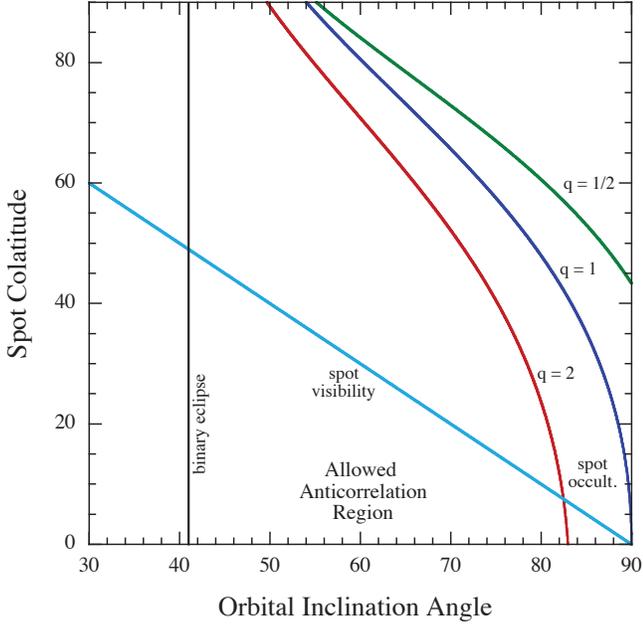}
\caption{Geometric requirements for the observer to be able to see a starspot around an entire orbit of an eclipsing binary. The allowed region lies below all the curves -- which is essentially dictated by the diagonal spot visibility line.  Eclipses will be seen for system parameters to the right of the vertical line, or even possibly above inclination angles as small as $34^\circ$. All angles are in degrees. The quantity $q \equiv M_2/M_1$.  The relative stellar radii are computed from $q$ which dictates the relative sizes of the two Roche lobes.}
\label{fig:visibility}
\end{center}
\end{figure}

An eclipse in a binary system consisting of two spherical stars of
radii $R_1$ and $R_2$ will only occur if the inclination satisfies
\begin{eqnarray}
i \gtrsim \cos^{-1} \left[\frac{R_1+R_2}{a}\right]
\label{eqn:occult}
\end{eqnarray} 
If we use the Eggleton (1983) analytic approximation for the size of
the Roche lobe for a range of mass ratios of $0.3 \lesssim M_2/M_1
\lesssim 3$, we find that $(R_1+R_2)/a$ is very close to 
$\sim$0.76, corresponding to a minimum inclination angle of
$\sim$$41^\circ$ (see Fig.~\ref{fig:visibility}).\footnote{For strictly 
contact binaries, the minimum inclination angle for eclipses to occur
is formally given as $34^\circ$ (see, e.g., Morris 1999).}

Finally, this model only produces strictly anticorrelated primary and secondary
$O-C$ curves if the spot is not occulted by the companion star.  If
the spot is not occulted when it is located at longitude $\ell = 0$,
then it will not be occulted when it is at any other
longitude\footnote{The proof of this is simple to visualize.  Project
the kinematic trajectory of the spot over the course of a binary orbit
onto a plane perpendicular to the line of sight and compare it to the
trajectory of the highest projected point on the companion star.}.  A
condition on the inclination angle to avoid occultation of the spot at
$\ell = 0$ is relatively straightforward to derive.  If the radii of
the star with the spot and the second star are $R_1$ and $R_2$
respectively, then the constraint can be written as one on the angle
$\alpha$ as a function of inclination:
\begin{equation} \label{eqn:i_requirement}
\alpha < \sin^{-1} \left[\frac{a}{R_1} \cos i - \frac{R_2}{R_1} \right] + i
\label{eqn:visible}
\end{equation}

The constraints on the inclination angle and the spot colatitude are
summarized in Fig.~\ref{fig:visibility}.  The spot colatitude is
plotted on the y axis and the inclination angle on the x axis.  Viable
regions in this parameter space lie to the right of the vertical line
at $41^\circ$, below the spot visibility line given by the requirement
that $\alpha + i < 90^\circ$, and below the curves given by
eq.~(\ref{eqn:visible}).

\begin{figure}
\begin{center}
\includegraphics[width=0.99 \columnwidth]{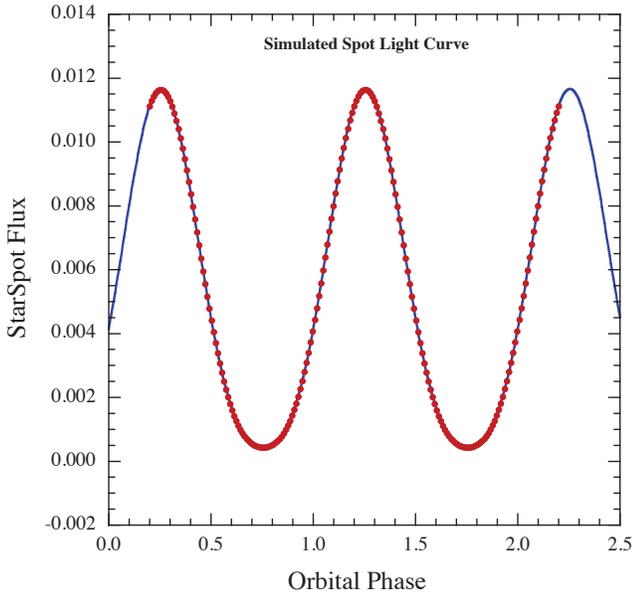}
\caption{{\em Phoebe} generated `light curve' of a single hot spot.  The red points are the relative flux coming from a hot spot as a function of the orbital phase.  To generate these points we adopted the following spot parameters: colatitude $\alpha = 45^\circ$, longitude $\ell = 90^\circ$, radius = 10$^\circ$, $T_{\rm spot}/T_{\rm star} = 1.15$.  The orbital inclination is $40^\circ$.  The blue curve is a model fit to equation (\ref{eqn:spotcurve}) in the text with $\ell$ set to 90$^{\circ}$.}
\label{fig:Phoebe_spot}
\end{center}
\end{figure}

\section{Light Curve Simulations with {\em Phoebe}}
\label{sec:Phoebe}

In order to verify the validity of some of the approximations used for
our simple model, we utilized the {\em Phoebe} binary light curve
emulator (Pr\v{s}a \& Zwitter 2005) to model a contact binary system
where either one cool or one hot spot may be present on one star. As a
strictly illustrative model, we utilized the {\em Phoebe} fit to the folded light
curve of KIC 3437800, a {\em Kepler} contact binary whose $O-C$ curves
are also anticorrelated, though it is not included in the present
sample of 32 systems.  That fit yielded the parameters $P_{\rm orb} =
8.7$ hours; $i = 40^\circ$, $T_{\rm eff} = 6185$ K, and $q = M_2/M_1 =
0.62$ that specify the baseline no-spot model.  A hot or cold spot was
then placed at one of a variety of locations on the surface of the
primary star, and orbital light curves were simulated.  The times of
the four extrema (two minima and two maxima) were
found using the same parabolic interpolation method used for the
actual {\em Kepler} data.  We emphasize that the system parameters
adopted for KIC 3437800 are illustrative only.  

For this particular example, the spot was positioned at $\alpha =
45^\circ$, $\ell = 90^\circ$, and was given a radius of 10$^\circ$
on the surface of the primary and a temperature that was elevated by
15\% over the local $T_{\rm eff}$ of the
star. Fig.~\ref{fig:Phoebe_spot} shows the difference between the
light curves of the models with and without the spot.  The difference
curve is not exactly a pure sine function.  This can readily be
understood in terms of limb-darkening, by modifying
eq.~(\ref{eqn:view_vector}) with a simple linear limb darkening law
such that:
\begin{eqnarray}
\Delta F_{\rm spot} \propto \cos \theta \left[1-u(1-\cos \theta)\right]
\end{eqnarray}
where $\cos \theta$ represents the dot product between the direction to the
observer and the spot vector with respect to the center of its host star, and
$u$ is the linear limb-darkening coefficient.  Substituting in the
expression for the dot product, we can write the above expression as:
\begin{eqnarray}
\label{eqn:spotcurve}
\Delta F_{\rm spot} &  =  & A + b(1-u+2au) \cos (\omega t +\ell) + ub^2  \cos^2 (\omega t +\ell) \nonumber \\  
 ~~~~a  & \equiv & \cos \alpha \cos i \nonumber ~~~~~~~{\rm and}~~~~~~~b  \equiv   \sin \alpha \sin i \\
\end{eqnarray}
where $A$ is a $DC$ offset. Basically, this is equivalent to equation (\ref{eqn:view_vector})
except for the addition of a $\cos^2$ term which accounts for the limb
darkening.  A best-fit curve of the form given by
eq.~(\ref{eqn:spotcurve}) is superposed on the data obtained from the
{\em Phoebe} simulations in Fig.~\ref{fig:Phoebe_spot}.  The fit is
excellent.

\begin{figure}
\begin{center}
\includegraphics[width=0.99 \columnwidth]{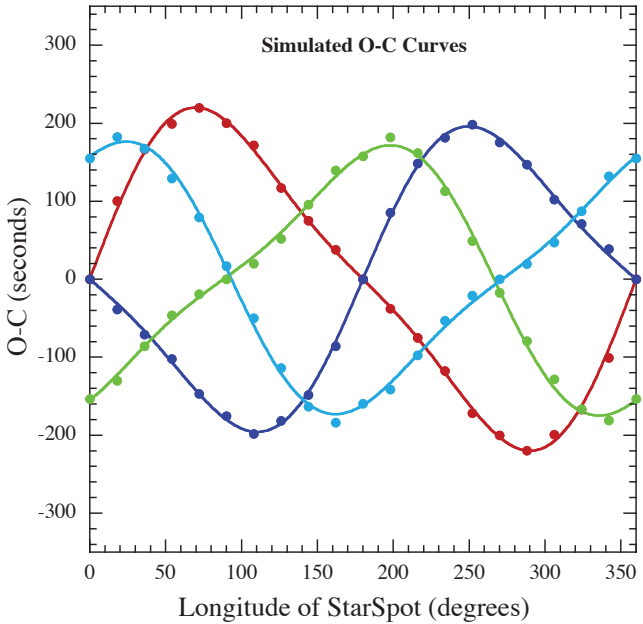}
\caption{$O-C$ curves generated by {\em Phoebe} for the binary system described in Fig.~\ref{fig:Phoebe_spot}, and modeled after KIC 3437800.  The red and blue points are for the primary and secondary eclipses, respectively, while the green and cyan points are for the out-of-eclipse maxima.  The smooth curves are fits to a simple spot model, modified by limb darkening, as discussed in Sect.~\ref{sec:Phoebe}.}
\label{fig:Phoebe_OmC}
\end{center}
\end{figure}

{\em Phoebe} model light curves were then computed for cases with the
spot centered at each of a set of longitudes covering the range
$0^\circ \leq \ell \leq 360^\circ$, and still at a colatitude of
$\alpha = 45^\circ$.  The $O-C$ results are shown in
Fig.~\ref{fig:Phoebe_OmC} for the two minima as well as the two
maxima.  It is immediately evident that the $O-C$ curves for the two
minima are indeed anticorrelated, as are the curves for the two
maxima; however, they are not pure sine curves.  The nonsinusoidal
behavior of the $O-C$ curves must differ from that in the simple model
because limb darkening produces a lightcurve for the spot that is not
exactly sinusoidal (see Fig.~\ref{fig:Phoebe_spot}).  When the
analytic expressions for the phase shifts, as given in
eqns.~(\ref{eqn:dphi}) and (\ref{eqn:dphix}), are rederived while
accounting for limb-darkening per eq.~(\ref{eqn:spotcurve}), highly
analogous results are found except that the $\sin \ell$ term in
eq.~(\ref{eqn:dphi}) is multiplied by a factor of $(1 \pm \chi \cos
\ell)$ while the $\cos \ell$ term in eq.~(\ref{eqn:dphix}) is
multiplied by a factor $(1 \pm \chi \sin \ell)$, where $\chi$ is a
geometry-dependent number of order $1/3-1/2$.  Therefore, even after
including limb darkening, the $O-C$ curves for the primary and
secondary minima (as well as for the two maxima) are still
anticorrelated in the sense that their values always have opposite
signs, however the magnitudes are not quite equal.

In Fig.~\ref{fig:Phoebe_OmC} we show fits to the {\em
Phoebe}-generated $O-C$ curves (filled circles) using the functions of $\ell$ 
given in eqns.~(\ref{eqn:dphi}) and (\ref{eqn:dphix}), but with each multiplied
by the extra factor discussed above.  The free parameter $\chi$ was
found to be consistent with 0.45 $\pm 0.04$ for all four curves.
Since the fits are quite good, it appears that the simple model
formalism captures the important aspects of the effects of starspots.

\section{Discussion}

The simple starspot model presented here to explain both the general
appearance and amplitude of the $O-C$ curves for contact binaries, as
well as the anticorrelations between the $O-C$ curves for the primary
and secondary eclipses, seems to work remarkably well.  At this point,
it is natural to wonder to what degree similar effects, especially the
anticorrelations between the $O-C$ curves for the primary and
secondary eclipses, would be observable in longer period binaries.

First, and perhaps most importantly, in order for the present starspot 
model to be effective in producing visibly detectable anticorrelated $O-C$ curves, 
the stars in the binary would be required to rotate nearly synchronously 
with the orbit.  With increasing orbital periods this becomes less and less
likely.  Second, for longer orbital periods, eclipses will only be seen in 
general for inclinations nearer to $90^\circ$.\footnote{Strictly speaking, eclipses
are not required for the spot model of anticorrelated $O-C$ curves to work;
however, wider, non-eclipsing binaries are more difficult to discover.}  
For larger inclination angles, 
starspots must be located nearer to the poles of the stars (see Fig.~{\ref{fig:visibility}) 
to avoid being occulted during the eclipses.  Not only will the unocculted 
region be smaller, but also spots may be less likely to occur near a stellar 
pole.  Third, it is plausible that contact binaries are more likely to have large 
spots and that any spots on them tend to be larger than the spots on the stars 
in longer period binaries.  

Finally, there will be a decrease in any $O-C$ amplitude due essentially to 
the smaller duty cycle of the eclipse in longer period binaries.  For longer 
orbital periods, the eclipse duration is given in terms of orbital cycles by:
\begin{eqnarray}
\frac{\Delta \theta_{\rm ecl}}{2 \pi} = \frac{1}{\pi} \sin^{-1}\left[\frac{(R_1+R_2)}{a}\right] \simeq \frac{(R_1+R_2)}{\pi a}
\end{eqnarray}
where the approximation has been made that $R_1 \sim R_2 \ll a$.  The
eclipse profiles may be crudely approximated as
in eq.~(\ref{eqn:binary_LC}) by taking $\mathcal{B} = -B \cos
N\omega t$ for times $t$ near those for which $\omega t = 0$ and $\omega t = \pi$.
Here $N$ would be given by
\begin{eqnarray}
N \simeq \frac{\pi}{\Delta \theta_{\rm ecl}} \simeq \frac{\pi a}{2(R_1+R_2)}
\label{eqn:N}
\end{eqnarray}
In the discussion presented above, e.g., in eq.~(\ref{eqn:binary_LC}),
we had taken $N=2$ to represent a contact binary.  For stars of a
fixed, typical, unevolved size, eq.~(\ref{eqn:N}) states that $N
\propto a \propto P_{\rm orb}^{2/3}$.  When the calculation that led
to eqs.~(\ref{eqn:dphi}) through (\ref{eqn:dtau}) is recast in terms
of $N$ with all other parameters held fixed, the leading factor of
$1/4$ becomes $1/N^2$.  In turn, this implies that
\begin{eqnarray}
 \tau \propto \frac{P_{\rm orb}}{N^2} \propto P_{\rm orb}^{-1/3}
\end{eqnarray}
This result may be applied, in an extreme example, to a binary with a
$\sim$30-day period, i.e., a period approximately two orders of
magnitude longer than that of a typical contact binary.  In a binary
with this longer period the spot-induced $O-C$ variations would be
reduced in amplitude by a factor of $\sim$5 over those of a contact
binary comprising similar stars and having similar spots.

We note that in keeping with these conclusions, based on the spot model,
close binaries are strongly favored to exhibit anticorrelated $O-C$
curves.  Table \ref{tab:stats} shows just how rapidly the anticorrelation 
phenomenon decreases with increasing orbital period.

Another question that arises in connection with the proposed starspot
model is what happens to the basic equation for shifts in the
timing, as in eq.~(\ref{eqn:dphi}), when there is more than one
starspot present at the same time?  The result is simply a sum of
terms as given in eq.~(\ref{eqn:dphi}) but with a distribution of
spot parameters, including different stellar longitudes -- the latter
being by far the most important.  For the case where a single spot of
a given area and temperature decrement is divided into $n$ smaller 
spots of the same {\em total} area, and assigned a random distribution 
of $\ell$ values, the net result will be a simple decrease in the amplitude 
of the resultant $O-C$ curve by roughly $\sqrt{n}$.  Thus, even if there are, 
e.g., 10 smaller spots present on one star, the amplitude of the shifts in 
timing are likely to be reduced by only a factor of a few.  If, by contrast, the 
number of spots increases to $n$, but their sizes and temperature decrements 
remain unchanged, then the amplitude of the $O-C$ curve {\em increases} by 
$\sqrt{n}$.  In either case, these collections of spots still have to migrate in a 
semi-coherent way, if quasi-periodic behaviors in the $O-C$ curves are to be 
observed.

\begin{figure}
\begin{center}
\includegraphics[width=0.99 \columnwidth]{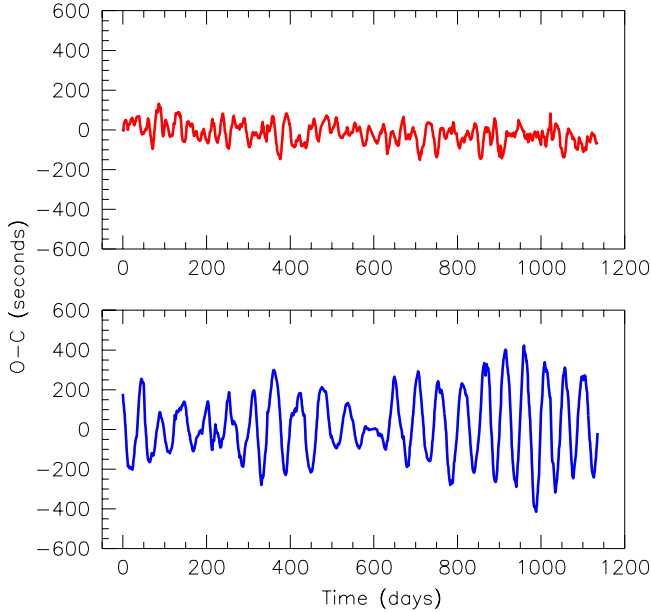}
\caption{ The sum (top) and difference (bottom) of the $O-C$ curves (divided by 2) 
for the primary and secondary eclipses of KIC 8956957, respectively.}
\label{fig:sumdiff_short}
\end{center}
\end{figure}

\begin{figure}
\begin{center}
\includegraphics[width=0.99 \columnwidth]{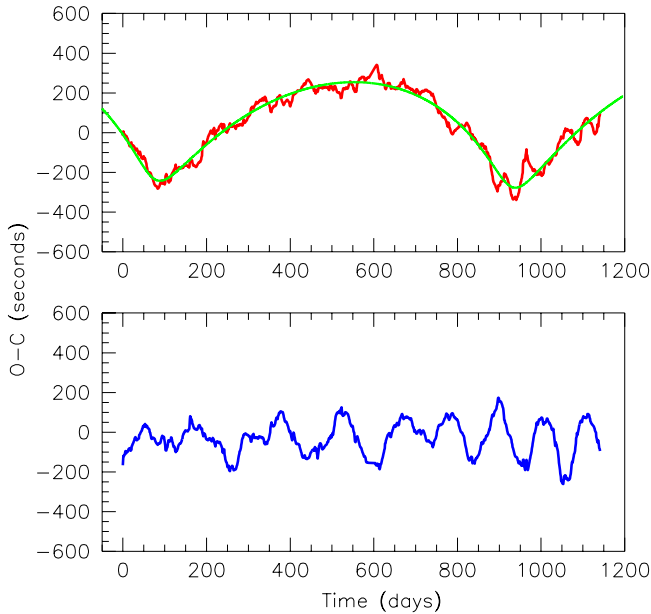}
\caption{ The sum (top) and difference (bottom) of the $O-C$ curves (divided by 2) 
for the primary and secondary eclipses of KIC 1873918, respectively. The solid 
green curve on the top panel represents an orbital fit to the $O-C$ (see text for details).}
\label{fig:sumdiff_long}
\end{center}
\end{figure}

The quasiperiodic $O-C$ variations with characteristic time scales
$\sim$50-200 days should carry information about the surface differential
rotation of the stellar components in our sample of contact binaries.
Following Kalimeris et al.~(2002) and Hall \& Busby (1990), we use a
differential rotation law of the form:
\begin{eqnarray}
P_{\alpha}  =  \frac{P_{\rm orb}}{(1- k \cos^2 \alpha)}
\end{eqnarray}
where, again, $\alpha$ is the colatitude of the spot, and $P_{\alpha}$
is the rotation period at colatitude $\alpha$.   The migration
period, $P_{\rm mig}$ is then:
\begin{eqnarray}
P_{\rm mig}  = \frac{P_{\rm orb}^2}{(P_{\alpha}-P_{\rm orb})} \simeq \frac{P_{\rm orb}}{ k \cos^2 \alpha}
\end{eqnarray}
where the right hand approximation is based on the assumption that $k
\ll 1$. If the 50-200 day time scales are interpreted as
migration periods, the values of $k$ must be in the range 0.003 -- 0.013, in
good agreement with, but covering a smaller range than, the values
cited by Kalimeris et al.~(2002) and Hall \& Busby (1990).
  
Finally, we noted above that the $O-C$ curves for some systems show
evidence of positive correlations between the primary and secondary
curves on relatively long time scales.  For contact binary systems
with a high degree of anticorrelated behavior between the $O-C$ curves
for the primary and secondary minima, there is a simple way to
separate out most of the starspot-induced eclipse timing changes from
most of the changes that represent other effects such as the
perturbations due to third bodies.  This involves forming the {\em
sum} of the two $O-C$ curves, divided by 2, as well as the {\em
difference}, divided by 2 (see also Conroy et al.~2013 who apply
the same technique).  The latter tends to emphasize the effects
of starspot activity, while the former tends to remove them, and
thereby possibly show more clearly any effects due to, e.g., a third
body. These sum and difference curves are particularly illuminating
for systems that show long timescale positively correlated $O-C$
curves, but with short timescale anticorrelated behavior.  This is
demonstrated in Fig.~\ref{fig:sumdiff_short} for KIC 8956957.  The top
panel shows the average of the two $O-C$ curves, and indicates very
little residual structure of interest.  By contrast, the bottom panel,
which shows the differences between the $O-C$ curves, clearly exhibits
the quasiperiodic behavior that we attribute to a starspot (or spotted
region).

Figure~\ref{fig:sumdiff_long} shows sum and difference $O-C$ curves
for KIC 1873918.  In this case, the summed curve (top panel) shows
long-term behavior that could be indicative of orbital motion induced
by a third star.  The superposed smoothed curve shows the results of
fitting for both the Roemer delay and the physical delay (see, e.g.,
Rappaport et al.~2013, and references therein; Conroy et al.~2013), both due to the
presence of a third star.  The inferred physical delay for this system
is very small, as expected from the short binary period and the much
longer inferred orbital period of the third star.  Aside from some
residual small-amplitude modulations in the $O-C$ curve, presumably
due to starspots, the fit is quite respectable.  The fitted parameters
are: $P_{\rm orb} \simeq 854$ days, eccentricity $\simeq 0.63$,
longitude of periastron $\omega \simeq 251^\circ$, Roemer amplitude
$\simeq 280$ s, and mass function $\simeq 0.032 \, M_\odot$.  This
object was missed as a candidate triple-star system in the initial
search by Rappaport et al.~(2013) likely because of the effects of
starspot activity.

Summed $O-C$ curves for all 32 of our illustrative sample of contact and ELV
binaries were computed and examined for possible evidence of a third
body.  The results for some six of the systems\footnote{These systems
include: KIC 2715007, 4937350, 7691553, 9020289, 9097798, and 9821923.}
show clear long term
quadratic trends which correspond to constant rates of change in the
orbital periods.  Typical values of the quadratic terms are $\dot
P_{\rm orb} \simeq 10^{-8}$ days/day, or $P_{\rm orb}/\dot P_{\rm orb}
\simeq 0.1$ Myr.  The presence of long term quadratic trends in the
$O-C$ curves of overcontact binaries is not unusual.  For example, Qian
(2001a; 2001b) lists 42 systems that evidently exhibit such features.
The mean value of $P_{\rm orb}/\dot P_{\rm orb}$ found by Qian
(2001a) for 12 contact binaries is 2.7 Myr, and for an additional
10 classified as `hot contact binaries' is 4.4 Myr.  All but one of
these has a positive sign, indicating a lengthening period.  Several
other systems are also listed elsewhere in the literature.  The Qian
(2001a) values for $\dot P_{\rm orb}$ thus represent a factor of $\sim$30
more slowly evolving periods than the handful that we are able to
detect in the {\em Kepler} collection.  The greater sensitivity of the
Qian (2001a) results is
due to the long historical baseline of the plates they utilized
($\sim$80 years vs. 4 years for the {\em Kepler} data base), in spite
of the lower precision in timing the eclipses.  We note, however, that
larger detected values of $\dot P_{\rm orb}$ in contact binaries are not 
unprecedented, e.g., LP UMa has $P_{\rm orb}/\dot P_{\rm orb}
\simeq 0.2$ Myr (Csizmadia, B\'{i}r\'o, \& Borkovits 2003).

In any case, such quadratic trends observed for the {\em Kepler} binaries 
might indicate the presence of a third body in an orbit with a period much 
longer than $\sim$1200 days, i.e., the length of the {\em Kepler} data train 
utilized in this work.  Or, such quadratic trends could possibly be explained by 
evolutionary effects, which manifest themselves via different forms of mass
exchange between the stellar components.  These could include mass transfer
in the context of thermal relaxation oscillation theory (Lucy 1976; Webbink 1976; 
Webbink 2003), or via angular momentum loss driven by stellar winds and/or
magnetic braking (see, e.g., van't Veer 1979; Mochnacki 1981; van't Veer \& 
Maceroni 1989).

\vspace{0.8cm}
\section{Summary and Conclusions}

Kalimeris et al. (2002) and earlier works showed that photometric
perturbations, and, in particular, starspots may affect measured $O-C$
curves, and that those perturbations are not properly interpreted in
terms of orbital period changes.  They noted that spot migration could
produce (quasi)periodic effects in the O-C curves.  

In this work, we have substantially extended these earlier results.  We have used the 
{\em Kepler} data base for binary stars, and an analytic model to provide good insight into
the timing effects of starspots seen in the $O-C$ curves.  In particular, we identified a large sample 
of {\em Kepler} target short-period binaries (i.e., $P_{\rm orb} \lesssim 1/2$ day) that appear to 
manifest the effects of a single spot or a small number of spots on their $O-C$ curves; these  
quite often have the form of a random-walk or quasiperiodic behavior, with typical amplitudes of 
$\sim$$\pm$ 300 s.  Most of these $O-C$ curves also exhibit a very pronounced anticorrelation 
between the primary and secondary minima.  

We developed 
a simple idealized model that illustrates the major effects that starspots have on measured 
eclipse times.  In particular we showed that a spot will, in general, affect the times of 
primary minimum and secondary minimum differently, with the predominant effect being 
an anticorrelated behavior between the two, provided that the spot is visible around much
of the binary orbit.  We also showed that a spot can equally well affect the times of the 
two maxima in each orbital cycle, and that the effects on the two maxima should be different, 
typically including anticorrelated behavior between them and a 90$^\circ$ phase shift with 
respect to the eclipse minima.

All of the same types of timing behavior are expected for close binaries that do not 
eclipse at all, i.e., so-called ``ELV'' binaries, and in fact, a significant fraction of the 
$\sim$390 binary systems exhibiting these properties may be in this category. There
is probably even a selection effect whereby the anticorrelation properties of the 
$O-C$ curves are enhanced in binaries that either do not eclipse or which have 
only partial eclipses.  The reason is that if an eclipse also occults a starspot over a 
substantial range of longitudes, $\ell$, on the surface, then the anticorrelation effect 
will be diminished.  We can even turn this argument around and suggest that the 
detection of anticorrelated $O-C$ curves tends to indicate that the system being 
observed is a binary, as opposed to a false positive, such as a pulsator.  This gives 
us some confidence that the four systems listed in Table \ref{tab:source} marked as false
positives by Matijevi\v{c} et al.(2012) are, in fact, actual binaries.

We have
found that a few of the selected contact binaries showed positively correlated variations in 
the $O-C$ curves for their primary and secondary minima on long time scales as well as the
anticorrelated variations that are most evident on shorter time scales.  We then demonstrated
that sum and difference $O-C$ curves between the primary and secondary eclipses are 
useful in distinguishing between the two types of variations.  We used this latter technique 
(i.e., of summing the two $O-C$ curves of the primary and secondary eclipses) to better isolate the 
effects of a possible third body in the system.  In the process we found a likely Roemer
delay curve for one of the systems, as well as convincing evidence for a long-term quadratic 
trend in six other systems. 

Finally, we found that the $O-C$ difference curves often 
appear to be dominated by 50 to 200 day quasiperiodicities that we interpret in terms of the
migration of spots, due to differential rotation, relative to the frame rotating with the orbital 
motion.

\acknowledgments
We thank A. Pr\v{s}a  for very helpful discussions.  The authors are grateful to the {\em Kepler} Eclipsing Binary Team for generating the catalog of eclipsing binaries utilized in this work.  We also thank Roberto Sanchis-Ojeda for his help in preparing the stitched data sets.  BK acknowledges support by the T\"UB\.ITAK (Project 112T776).  The project has been partially supported by the City of Szombathely under agreement No. S-11-1027.

\input{source.tbl}

\end{document}

%% file: source.tbl.tex
\begin{deluxetable}{lccccccc}
\tablewidth{0pt}
\tabletypesize{\scriptsize}
\tablecaption{\label{tab:source} {\em Kepler} Contact Binaries with Anticorrelated $O-C$ Curves}
\tablehead{
	\colhead{Source} &
	\colhead{Binary Period} &
	\colhead{Morphology$^{1}$}  & 
	\colhead{Magnitude} &
	\colhead{$T_{\rm eff}$} &
    	\colhead{Correlation} &
	\colhead{Minimum 1} &
	\colhead{Minimum 2} \\
	\colhead{KIC \#} & 
	\colhead{(days)} &
	\colhead{$(0-1)$} & 
      	\colhead{($K_p$)} &
	\colhead{(K)} &
	\colhead{Coefficient$^2$} &
	\colhead{Depth$^3$} &
	\colhead{Depth$^3$}
}
\startdata 
1873918$^4$	& 0.332433 & 0.86 & 13.72 & 5715 & 0.42$^{(5)}$ & 0.104 & 0.099\\
2017803$^4$	& 0.305742 & 0.97 & 14.61 & 5056 & -0.50 & 0.046 & 0.036 \\
2159783	& 0.373886 & 0.87 & 14.96 & 5643 & -0.16 & 0.217 & 0.194 \\
2715007	& 0.297107 & 0.87 & 14.73 & 5598 & 0.10  & 0.025 & 0.018 \\
3837677	& 0.461984 & 0.94 & 15.55 & 5466 & -0.46 & 0.104& 0.096 \\
3853259	& 0.276648 & 1.00 & 13.92 & 4467 & -0.42  & 0.076 & 0.073 \\
4937350$^4$	& 0.393664 & -1.00 & 14.27 & 5862 & -0.60 & 0.072 & 0.068 \\ 
5008287	& 0.291878 & 0.93 & 15.31 & 5881 & -0.34 & 0.050 & 0.048 \\
5022573	& 0.441724 & 0.98 & 11.47 & 5648 & -0.13 & 0.056 & 0.056 \\
5033682	& 0.379916 & 0.95 & 13.26 & 5611 & -0.63 & 0.056 & 0.034 \\
5283839	& 0.315231 & 0.92 & 15.16 & 5906 & 0.09  & 0.160 & 0.146 \\
6964796	& 0.399966 & 0.97 & 12.61 & 5657 & -0.52 & 0.050 & 0.044 \\
7118656	& 0.321355 & 0.94 & 15.03 & 5271 & -0.24 & 0.052 & 0.040 \\
7217866	& 0.407157 & 0.90 & 13.86 & 5600 & -0.15  & 0.106 & 0.093 \\
7542091	& 0.390499 & 0.81 & 12.34 & 5673 & -0.31 & 0.154 & 0.143 \\
7691553	& 0.348309 & 0.93 & 14.62 & 5786 & -0.48  & 0.092 & 0.088 \\
7773380	& 0.307577 & 0.94 & 14.47 & 5357 & -0.40  & 0.084 & 0.066 \\
8190613	& 0.332584 & 0.90 & 15.13 & 5384 & -0.28 & 0.117 &  0.113 \\
8956957	& 0.324382 & 0.96 & 13.98 & 6307 & -0.72  & 0.056 & 0.054 \\
9020289	& 0.384027 & 0.94 & 14.74 & 5997 & -0.60  & 0.061 & 0.059 \\
9071104	& 0.385213 & 0.81 & 13.65 & 5959 & -0.52  & 0.158 & 0.130 \\
9097798	& 0.334068 & 1.00 & 14.58 & 5592 & -0.41 & 0.027 & 0.026 \\
9451598	& 0.362349 & 0.93 & 13.63 & 6060 & -0.77  & 0.041& 0.030 \\
9519590	& 0.330895 & 0.88 & 13.92 & 5961 & -0.24  & 0.039 & 0.036 \\
9821923	& 0.349532 & 0.95 & 14.21 & 5730 & -0.53  & 0.087 & 0.083 \\
9832227	& 0.457950 & 0.94 & 12.26 & 5854 & -0.19  & 0.094 & 0.086 \\
10148799 & 0.346605 & 0.91 & 15.41 & 5340 & -0.66 & 0.068 & 0.037 \\
10155563 & 0.360268 & 0.94 & 11.99 & 5982 & - 0.03 & 0.017 & 0.013 \\
11013608$^4$ & 0.318287 & -1.00 & 12.57 & 6223 & -0.33 & 0.048 & 0.044 \\
12055421 & 0.385607 & 0.96 & 12.52 & 6118 & -0.47 & 0.041 & 0.038 \\
12418274 & 0.352723 & 0.93 & 14.35 & 5215 & 0.22  & 0.050 & 0.038 \\
12598713 & 0.257179 & 0.94 & 12.76 & 5189 & -0.14  & 0.013 & 0.008 \\ 
\enddata
\tablecomments{All parameters, unless otherwise specified, are from \url{http://keplerebs.villanova.edu/} (Matijevi\v{c} et al.~2013). (1) The binary light curve morphology index is defined such that detached systems have values $\lesssim 0.5$; semi-detached systems are in the range $0.5-0.7$; and overcontact binaries lie in the range $0.7-0.8$.  Systems with morphology index above $\sim$0.8 are ellipsoidal variables and or uncertain categories.  Systems with -1 are unclassified. (2) A description of how the correlation coefficients were computed is given in Sect.~\ref{sec:CBOmC}.  (3) Depths of the primary and secondary minima in our folded light curves.  (4) These systems are labeled as ``binary false positives'' by the {\em Kepler} team.  We believe that the anticorrelated light curves provide evidence that they are, in fact, binaries.  (5) The significant positive correlation for this system arises from the likely Roemer delay in the $O-C$ curve due to the possible presence of a third body.}
\end{deluxetable}